# LHC Databases on the Grid: Achievements and Open Issues[*]

A.V. Vaniachine

*Argonne National Laboratory*
*9700 S Cass Ave, Argonne, IL, 60439, USA*

**Abstract:** To extract physics results from the recorded data, the LHC experiments are using Grid computing infrastructure. The event data processing on the Grid requires scalable access to non-event data (detector conditions, calibrations, etc.) stored in relational databases. The database-resident data are critical for the event data reconstruction processing steps and often required for physics analysis.

This paper reviews LHC experience with database technologies for the Grid computing. List of topics includes: database integration with Grid computing models of the LHC experiments; choice of database technologies; examples of database interfaces; distributed database applications (data complexity, update frequency, data volumes and access patterns); scalability of database access in the Grid computing environment of the LHC experiments. The review describes areas in which substantial progress was made and remaining open issues.

## 1. Introduction

In 2010 four experiments at the Large Hadron Collider (LHC) started taking valuable data in the new record energy regime. In preparations for data taking, the LHC experiments developed comprehensive distributed computing infrastructures, which includes numerous databases. This paper reviews LHC experience with database technologies for the Grid computing, including areas in which substantial progress was made. I was responsible for Database Deployment and Operations (formerly called Distributed Database Services) in the ATLAS experiment since 2004. As a result, this review is biased towards my personal views and experience. Beyond ATLAS, I started compiling information on databases in LHC experiments for my earlier invited talks on the subject [1, 2].

As an example of what relational databases used for in each LHC experiments, I briefly describe ATLAS database applications. In ATLAS, there are more than fifty database applications that reside on the central ("offline") Oracle server. By February 2010 ATLAS accumulated more than 8 TB of data, which are dominated by 4 TB of slow control data. Most of these database applications are "central" by their nature, like the ATLAS Authorship Database used to generate author lists for publications. The central databases are traditional applications developed according to standard Oracle best practices with a help of our OCP database administrators. Because these database applications are designed by traditional means, I will not cover LHC experience with these central applications in this review, since I cannot claim that LHC advanced the existing knowledge base in these traditional areas.

---

[*] Invited talk presented at the IV International Conference on "Distributed computing and Grid-technologies in science and education" (Grid2010), JINR, Dubna, Russia, 28 June - 3 July, 2010.

In contrast to the central database applications that are accessed by people or by limited number of computers and do not have to be distributed, a subset of LHC database applications must be distributed worldwide (for scalability) since they are accessed by numerous computers (Worker Nodes) on the Grid.

## 2. Database Applications and Computing Models of LHC Experiments

The LHC experiments are facing an unprecedented multi-petabyte data processing task. To address that challenge LHC computing models adopted Grid computing technologies. These computing models of LHC experiments determined the need for distributed database access on the Grid. The LHC Computing models are well represented at this conference in several talks and reviews [3, 4]. In essence, the LHC computing models are mostly focused on the problem of managing the petabyte-scale event data that are kept in a file-based data store, with files catalogued in various databases. These event store databases are an integral part of the LHC computing models. A brief description of a growing LHC experience with the event store database as it approach petasacles is provided in the last section.

In addition to the file-based event data, LHC data processing and analysis require access to large amounts of the non-event data (detector conditions, calibrations, etc.) stored in relational databases. In contrast to the file-based LHC event store databases, the database-resident data flow is not detailed in the "big picture" of LHC computing models. However, in this particular area the LHC experiments made a substantial progress compared with other scientific disciplines that use Grids. That is why I will focus on the LHC experience with distributed database applications introduced in the next section.

## 3. Distributed Database Applications Overview

In ATLAS there are only few database applications that have to be distributed: Trigger DB, Geometry DB, Conditions DB and Tag DB. ATLAS developed the Trigger DB for central ("online") operations. A small subset of the whole database is distributed on the Grid in SQLite files for use in Monte Carlo simulations. To manage the detector description constants ("primary numbers") ATLAS developed the Geometry DB with contributions from LHC Computing Grid (LCG). It is the first ATLAS database application that was deployed worldwide. It is distributed on the Grid in SQLite replica files. The Conditions DB was developed by LCG with ATLAS contributions. The LCG technology for Conditions DB is called COOL. The Conditions DB is a most challenging database application. It is a hybrid application that includes data in RDBMS and in files. Conditions data are distributed worldwide via Oracle Streams and via files. The ATLAS Tag DB stores event-level metadata for physics (and detector commissioning). It was developed by LCG with ATLAS contributions. It is distributed worldwide in files and also 4 TB are hosted at select Oracle servers. The Tag DB is expected to collect 40 TB of data per nominal LHC year of operations. Given the limited data taken to date, we have not yet gathered much experience in large-scale Tag DB access.

Another LHC experiment that adopted common LCG technology for Conditions DB—COOL (Conditions database Of Objects for LHC)—is LHCb. In COOL database application architecture the Interval-of-Validity (IOV) metadata and a data payload, with an optional version tag, usually characterize the conditions data. Similar Conditions database architecture was developed by the CMS experiment (Fig. 1). The CMS conditions database stores time-varying data (calibration and alignment) together with their sequences versions IOV. IOV sequences are identified by tags, which are organized as trees. A tag tree can be traversed from any node (global tag).

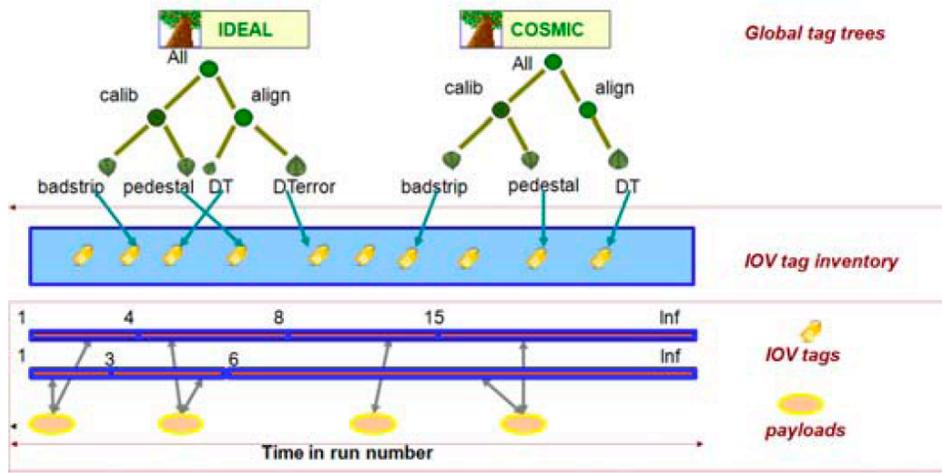

**Figure 1:** The CMS conditions database architecture.

As in ATLAS, the ALICE Conditions DB is also hybrid, comprised of data stored in the Offline Conditions Database (OCDB) a set of entries in the AliEn file catalogue pointing to files stored in the Grid. Together with the conditions data, the reference data are also stored in an analogous database.

## 4. Distributed Database Applications Details

This section describes database applications requirements, data complexity, update frequency, data volumes, usage, etc. Generally, these factors dictate the choice of database technologies: relational or hybrid (chosen in ATLAS and ALICE experiments for Conditions DB implementation).

*4.1. ATLAS Geometry DB*

Due to differences in requirements and implementation, ATLAS Geometry DB is separated from the Conditions DB to keep static information, such as nominal geometry. Only the time-dependent alignment corrections to Geometry are stored in the Conditions DB. Such separation of concerns resulted in a moderate data complexity of the Geometry DB. A recent statistics (February, 2010) counted 434 data structures described in 872 tables in one schema. The total number of rows was 555,162, resulting in an SQLite replica volume of 33 MB. The update frequency of the Geometry DB is "static" i.e. upon request, when the geometry corrections or updates become necessary. The database is accessed via a low-level common LCG database access interface called Common Object-Relational Access Layer (CORAL).

A typical data reconstruction job makes about 3K queries to the Geometry database. The master Geometry DB resides in the "offline" Oracle, where it is not used for production access. For example, for the Tier-0 operations an SQLite snapshot replica is made nightly. The Geometry DB is replicated on the Grid via SQLite files. During 2009 twenty-nine SQLite snapshots were distributed on the Grid, in 2008 it was eighteen.

*4.2. ATLAS Conditions DB*

Driven by the complexity of the subdetectors requirements, ATLAS Conditions DB technology is hybrid: it has both database-resident information and external data in separate files, which are referenced by the database-resident data. These external files are in a common LHC format called POOL. ATLAS database-resident information exists in its entirety in Oracle but can be distributed in smaller "slices" of data using SQLite. Since Oracle was chosen as a database technology for the "online" DB, ATLAS benefits of uniform Oracle technology deployment down to the Tier-1 centers. Adoption of Oracle avoids translating from one technology to another and leverages Oracle support from CERN IT and WLCG 3D Services [6].

Historically, ATLAS separated conditions database instances for Monte Carlo simulations and for the real data. The two instances still remain separate to prevent accidental overwrite of the Conditions DB for real data. Both Conditions DB instances are accessed via common LCG interface COOL/CORAL. This approach is similar to the CMS Conditions DB partitioning by usage (see below).

The complexity of the ATLAS Conditions DB data for simulations is high. According to a representative snapshot of February, 2010 the instance has 2,893 tables in four schemas. The total number of rows is 842,079 and the data volume of the SQLite replica is 376 MB. There are additionally 247 MB of data in 1049 POOL/ROOT files grouped in 25 datasets. The update frequency is "static," i.e. the database is updated upon request typically in preparation for major Monte Carlo simulations campaigns. All conditions data for Monte Carlo simulations is replicated on the Grid vial files (the full snapshot in SQLite plus the external POOL/ROOT files and their catalogs.).

The ATLAS Conditions DB for real data has a very high complexity. In February 2010, the database had 7,954 tables in 29 active schemas out of 45 schemas total. The schema count is determined by the number of ATLAS detector subsystems: 15 subsystems each having two schemas ("online" and "offline") plus one inactive combined schema (to be decommissioned). . The total number of rows is 761,845,364 and the Oracle data volume is 0.5 TB. There are additionally 0.2 TB in POOL/ROOT files grouped in 48 datasets. The Conditions DB for real data is updated continuously. Because of the large volume, use of the full database replica on the Grid is not practical. Only the required "slices" of the ATLAS Conditions DB data are distributed on the Grid. To process a 2 GB file with 1K raw events a typical reconstruction job makes about 11K queries to read more than 70 MB of database-resident data (with some jobs read tens of MB extra) plus about ten times more volume of data is read from the external POOL files.

*4.2. LHCb Conditions DB*

The LHCb reconstruction and analysis jobs are making direct connection via COOL/CORAL libraries from the Worker Nodes on the Grid to the Oracle replicas at the Tier-1 sites. Jobs require a limited amount of data transfer (~40 MB) in the first few minutes. SQLite replicas are used in the used in special cases, such as Monte Carlo simulations.

*4.3. ALICE Conditions DB*

Figure 2 shows conditions data flow in Shuttle—a special service providing an interface between the protected online world and the external computing resources [5].

Since 2008 the ALICE Conditions DB accumulated more that 30 GB of data for about 183,000 files plus more than 8 GB of the reference data for more than 29,000 files. All collected conditions data are exported on the Grid, thus making them accessible for the reconstruction and analysis.

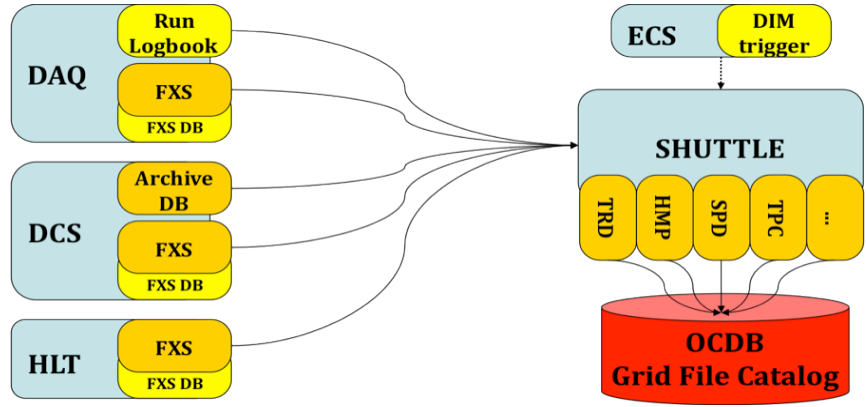

**Figure 2:** ALICE Shuttle framework for Offline Conditions DB (OCDB) [5].

*4.4. CMS Conditions DB*

All relations in Conditions DB are purely logical and application specific. As in case of other LHC experiments, no RDBMS consistency enforced, which allows full flexibility in copying (deep and shallow) at all level of the structure. As in case of ATLAS, data consistency is enforced not by database design but through a set of policies, such as NO DELETE, NO UPDATE. In CMS, only the current IOV sequence can be extended. In contrast to ATLAS, the data payloads are implemented as POOL/ORA objects stored in the database internally.

As in ATLAS, the CMS Conditions DB has been "partitioned" into schemas following development and deployment criteria, which keep separated areas of independent development: by sub-detectors, by software release. These are "partitioned" further by use-cases to keep separated independent workflows use cases. In case of Monte Carlo simulations, all relevant data are copied into a dedicated schema including even a single SQLite file. In case of re-processing at remote Tier-1 sites, a read-only snapshot of the whole Conditions DB is made for access through Frontier. Making the replica copy read-only prevents accidental overwrites, since the master Conditions DB is continuously updated for use in prompt data processing: reconstruction, calibration, and analysis at Tier-0. The database is managed through application-specific tools described in the next section. A CMS data reconstruction job reads about 60 MB of data.

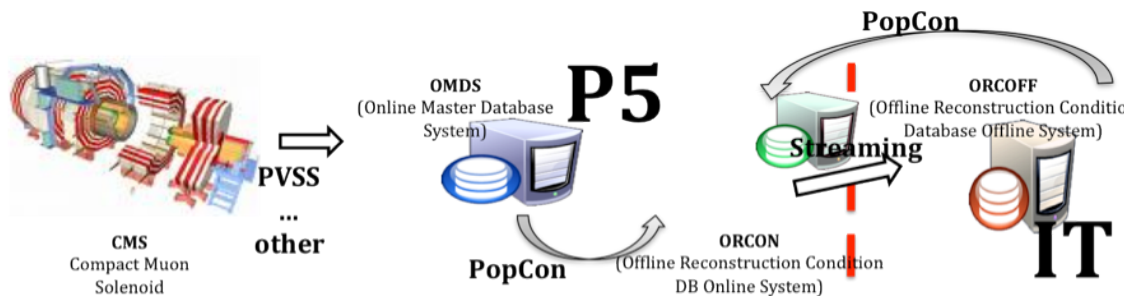

**Figure 3:** CMS database tool PopCon is used in all three CMS Oracle databases for the conditions data.

## 5. Database Integration

This section provide examples of custom database interfaces and tools on top of CORAL/COOL and describes integration of databases with software frameworks and into an overall data acquisition, data processing and analysis chains of the experiments.

Figure 3 presents an example of a database tool is the CMS PopCon (Populator of Condition objects). Fully integrated in the overall CMS framework, PopCon is an application package intended to transfer, store, and retrieve condition data in the "offline" databases. PopCon also assigns metadata information: tag and IOV.

Support for on-demand data access—a key feature of the common Gaudi/Athena framework—emphasizes the importance of database interfaces for LHCb and ATLAS experiments. On-demand data access architecture makes Oracle use straightforward. In contrast, the delivery of the required Conditions DB data in files is challenging, but can be implemented for a well-organized workflow, such as reprocessing. In the on-demand data access environment having a redundant infrastructure for database access turns out to be advantageous. The redundancy is achieved through common LHC interfaces for persistent data access, which assure independence on available technologies (Oracle, SQLite, Frontier…). No changes in the application code are needed to switch between various database technologies (Fig. 4). In ATLAS, each major use case is functionally covered by more than one of the available technologies, so that we can achieve a redundant and robust database access system.

In addition, various tools can be built on top of the interfaces. For example, since the large volume of ATLAS Conditions DB prevents use of the full database replica on the

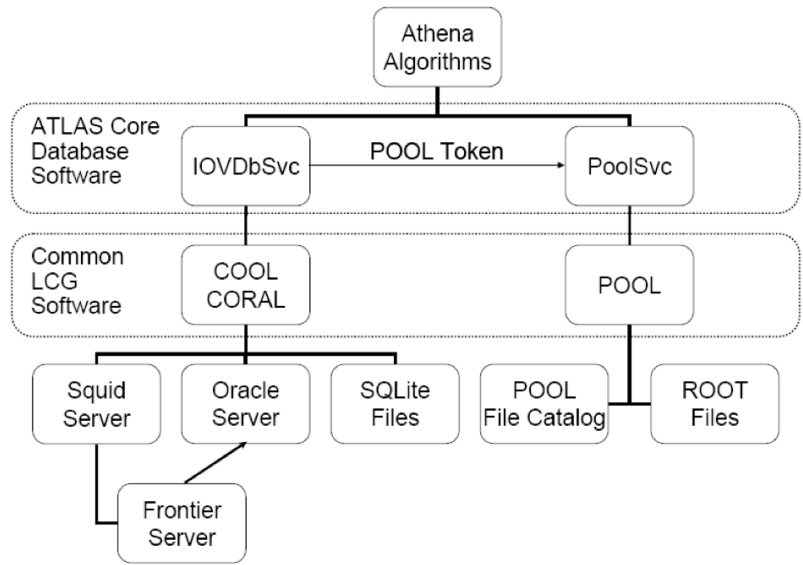

**Figure 4:** Integration of common LCG interfaces for database access in case of the ATLAS software framework Athena.

Grid, an advanced "db-on-demand" tool was developed to produce "slices" of the required conditions data for the Grid jobs [7].

## 6. Scalability of Database Access on the Grid

Scalability of database access in the distributed computing environment is a challenging area in which a substantial progresses was made by the LHC experiments.

### 6.1. ATLAS Database Release Technology

In a non-Grid environment, in case of ATLAS, two solutions assure scalability of access to Conditions DB database: a highly replicated AFS volume for the Conditions POOL files and the throttling of job submission at Tier-0 batch system. None of Tier-0 solutions for scalable database access is available on the Grid. As a result, ATLAS experience with database access on the Grid provided many useful "lessons learned."

In 2004, we found that the chaotic nature of Grid computing increases fluctuations in database load: daily fluctuations in the load are fourteen times higher than purely statistical [8]. To avoid bottlenecks in production, the database servers capacities should be adequate for a peak demand [6]. In 2005, to overcome scalability limitations in database access on the Grid, ATLAS introduced the Database Release concept [9]. Conceptulally similar to the software release packaging for distribution on the Grid, the Database Release integrates all necessary data in a single tar file:
- the Geometry DB snapshot as an SQLite file,
- a full snapshot of Conditions DB data for Monte Carlo in the SQLite file,
- plus corresponding Conditions DB POOL files and their POOL File Catalogue.

Years of experience resulted in continuous improvements in the Database Release approach, which now provides solid foundation for ATLAS Monte Carlo simulation in production [10]. In 2007 the Database Release approach was proposed as a backup for database access in reprocessing at Tier-1 sites (Fig. 5).

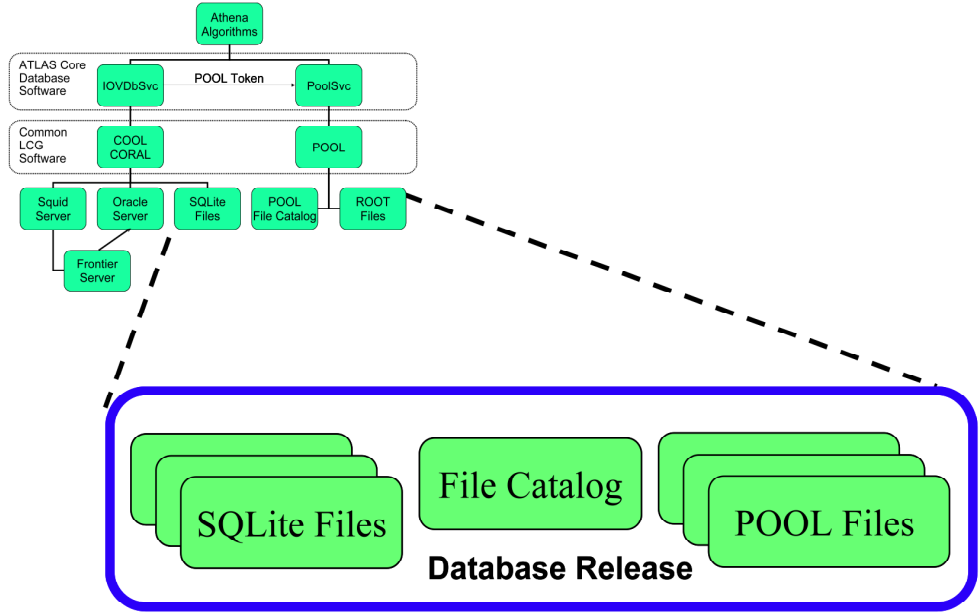

**Figure 5:** ATLAS Database Release technology hides the complexity of Conditions DB access (Fig. 4).

In addition to Database Releases, ATLAS Conditions DB data are delivered to all ten Tier-1 sites via continuous updates using Oracle Streams technology [11]. To assure scalable database access during reprocessing ATLAS conducted Oracle stress-testing at the Tier-1 sites. As a result of stress-tests we realized that the original model, where reprocessing jobs would run only at Tier-1 sites and access directly their Oracle servers, would cause unnecessary restrictions to the reprocessing throughput and most likely overload all Oracle servers [12].

Thus, the DB Release approach, developed as a backup, was selected as a baseline. The following strategic decisions for database access in reprocessing were made:
- read most of database-resident data from SQLite,
- optimize SQLite access and reduce volume of SQLite replicas,
- maintain access to Oracle (to assure a working backup technology, when required).

As a result of these decisions ATLAS DB Release technology fully satisfies the Computing Model requirements of data reprocessing and Monte Carlo production: it is fast (less than 10 s per job), robust (failure rate less than $10^{-6}$ per job) and scalable: (served ~1B queries in one of reprocessing campaigns). The read-only Database Release dataset guarantees reproducibility and prevents access to unnecessary data (similar to CMS partitioning by usage).

*6.2. CMS Frontier/Squid Technology*

Frontier/Squid is a data caching system providing advantages for distributed computing. To achieve scalability, the system deploys multiple layers of hardware and software between a database server and a client: the Frontier Java servlet running within a Tomcat servlet container and the Squid—a single-threaded http proxy/caching server (Fig. 6) [13]. Depending on a fraction of the shared data required by the jobs, the time needed to retrieve conditions data at the beginning of a job is reduced by factors of 3 to 100, depending on the distance between the site running the job and the remote site providing the Oracle conditions database.

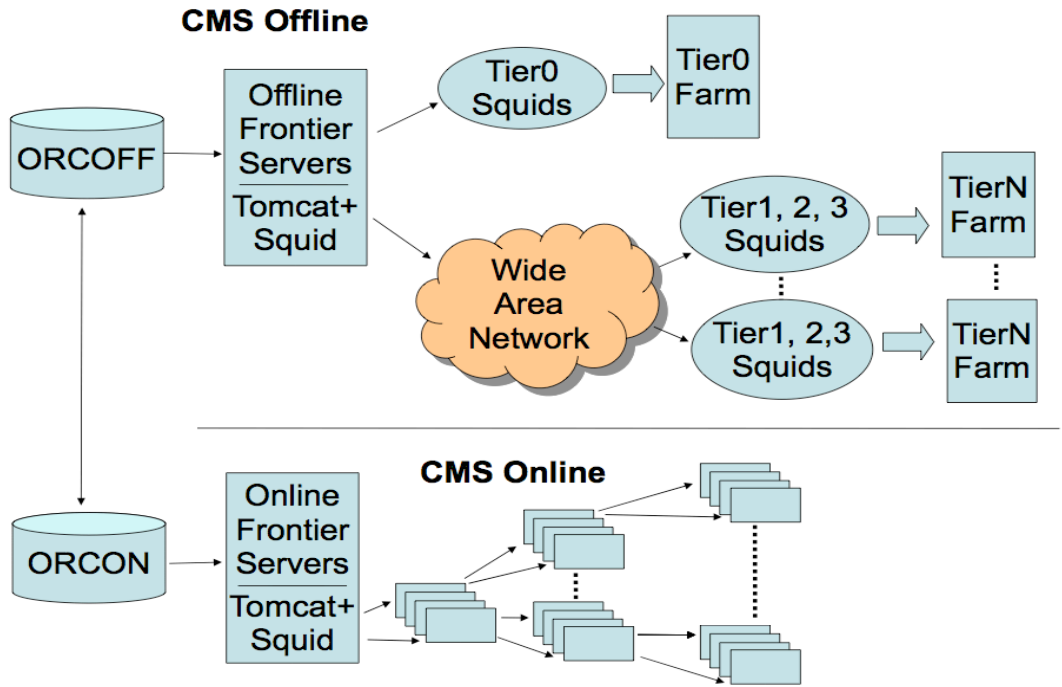

**Figure 6:** CMS Frontier/Squid deployment architecture.

To reduce a chaotic load on the Oracle databases at the Tier-1 sites caused by the the analysis jobs, ATLAS adopted the CMS Frontier/Squid technology, which have been shown to drastically reduce this load. This greatly improves the robustness of the conditions data distribution system. Having multiple Frontier servers has provided redundancy. For that, ATLAS has implemented Frontier servers at most of the Tier-1 sites and Squid servers at all Tier-0/1/2. Work is underway to provide Squid servers at most ATLAS Tier-3 sites.

**7. Scalability of the LHC Event Store Database**

Due to a hardware errors data corruption is inevitable in a large-scale data store. Event on a smaller scales, a single bit-flip in a file result in a corruption of the whole file unless some dedicated data recovery techniques are in use. Similarly, in case of the LHC event store we must discard a whole dataset with thousands of files if a single file is corrupted. However, at a certain data corruption rate this approach is not scalable, since a very large dataset will waste a lot of attempts during production. To assure event store scalability, LHC experiments introduced redundant higher-levels checksums to detect these types of errors. In ATLAS, every event store file is checked immediately after it was produced. The check verifies that the POOL/ROOT zlib compressed data buffers have correct checksums. If the file is unreadable the job marked as failed and re-executed.

LHC experience shows that we must introduce considerable redundancy, in order to detect and recover from data corruption errors, since these errors are costly to fix. The next redundant check is done at the end of each Grid job, when the checksum is calculated for each file produced at the Worker Node. This checksum is compared with the checksum calculated for the file transferred to the Storage Element by the LHC data transfer tools. In case of the checksum mismatch, the job is marked as failed and re-executed. Sites that did not implemented this check produce silent data corruption, where the mismatch is discovered at a later stage. This is not scalable, since, correcting silent data corruption in a distributed petascale event store is very costly. To assure scalability, the data corruption must be detected at the spot.

Learning from the initial operational experience, LHC experiments realized that the end-to-end data integrity strategies need to be developed for petascale data store. In a petascale event store, every layer of services should not assume that the underlying layer never provide corrupted or inconsistent data.

**8. Summary**

LHC experiments developed and deployed distributed database infrastructure ready for the LHC long run In ATLAS each major use case is functionally covered by more than one of the available technologies to assure a redundant and robust database access In CMS a novel http data caching system—Frontier/Squid—assures scalability of Conditions DB access on the Grid. The Frontier/Squid technologies are adopted by ATLAS as a part of end-to-end solution for Conditions DB access in user analysis. These technologies for distributed database access represents the area where the LHC experiments made a substantial progress, compared with other scientific disciplines that use Grids. Remaining open issues provide roadmap for future R&D in the area of scalable tools for data-intensive scientific discovery.

**Acknowledgements**

I wish to thank the Conference organizers for their invitation and hospitality. I also thank all my collaborators and colleagues who provided materials for this review. This work supported in part by the U.S. Department of Energy, Division of High Energy Physics, under Contract DE-AC02-06CH11357.